\documentstyle[12pt]{article}
\renewcommand{\r}{\rho} 
\advance\voffset by 15mm   
\makeatletter
\@addtoreset{equation}{section}
\makeatother

\def\ii{\'\i}

\def\ftoday{{\sl {Le \number\day \space\ifcase\month 
\or janvier\or f\'evrier\or mars\or avril\or mai
\or juin\or juillet\or ao\^ut\or septembre\or octobre
\or novembre \or d\'ecembre\fi\space \number\year}}}    
\def\ptoday{{\sl {\number\day \space de\space \ifcase\month 
\or janeiro\or fevereiro\or mar{\c c}o\or abril\or maio
\or junho\or julho\or agosto\or setembro\or outubro
\or novembro \or dezembro\fi\space de\space \number\year}}}    
\def\gtoday{{\sl {Den \number\day. \ifcase\month 
\or Januar\or Februar\or M\"arz\or April\or Mai
\or Juni\or Juli\or August\or September\or Oktober
\or November \or Dezember\fi\space \number\year}}}    
\def\today{{\sl {\ifcase\month
\or January\or February\or March\or April\or May
\or June\or July\or August\or September\or October
\or November \or December\fi \space\number\day,\space 
                                            \number\year}}}

\renewcommand{\a}{\alpha}
\renewcommand{\b}{\beta}
\newcommand{\g}{\gamma}           \newcommand{\G}{\Gamma}
\renewcommand{\d}{\delta}         \newcommand{\D}{\Delta}
\newcommand{\e}{\varepsilon}

\newcommand{\la}{\lambda}        
\newcommand{\m}{\mu}
\newcommand{\n}{\nu}
         
\newcommand{\p}{\psi}              
\newcommand{\s}{\sigma}           \renewcommand{\S}{\Sigma}
         
\newcommand{\f}{{\phi}}           \newcommand{\F}{{\Phi}}
\newcommand{\vf}{{\varphi}}

\newcommand{\SS}{{\cal S}}

\newcommand{\es}{\\[3mm]}

\newcommand{\sla}{\raise.15ex\hbox{$/$}\kern -.57em} 
\newcommand{\Sla}{\raise.15ex\hbox{$/$}\kern -.70em}
\def\h{\hbar}

\newcommand{\complex}{{\kern .1em {\raise .47ex
\hbox {$\scriptscriptstyle |$}}
    \kern -.4em {\rm C}}}
\newcommand{\real}{{{\rm I} \kern -.19em {\rm R}}}
\newcommand{\rational}{{\kern .1em {\raise .47ex
\hbox{$\scripscriptstyle |$}}
    \kern -.35em {\rm Q}}}
\renewcommand{\natural}{{\vrule height 1.6ex width
.05em depth 0ex \kern -.35em {\rm N}}}

\newcommand{\pa}{\partial}

\newcommand{\dfud}[2]{{\displaystyle{\frac{\delta #1}{\delta #2}}}}
\newcommand{\dfrac}[2]{{\displaystyle{\frac{#1}{#2}}}}

\newcommand{\cf}{{\em cf.\ }}
\newcommand{\twiddle}{\lower.9ex\rlap{$\kern -.1em\scriptstyle\sim$}}

\newcommand{\equ}[1]{(\ref{#1})}
\newcommand{\eq}{\begin{equation}}
\newcommand{\eqn}[1]{\label{#1}\end{equation}}
\newcommand{\eea}{\end{eqnarray}}
\newcommand{\eqa}{\begin{eqnarray}}
\newcommand{\eqan}[1]{\label{#1}\end{eqnarray}}
\newcommand{\ba}{\begin{array}}
\newcommand{\ea}{\end{array}}
\newcommand{\eqac}{\begin{equation}\begin{array}{rcl}}
\newcommand{\eqacn}[1]{\end{array}\label{#1}\end{equation}}

\advance\voffset by -2cm

\def\a{\alpha}
\def\b{\beta}
\def\d{\delta}
\def\e{\epsilon}

\def\f{\phi}
\def\vf{\varphi}
\def\g{\gamma}
\def\h{\eta}

\def\l{\lambda}
\def\m{\mu}
\def\n{\nu}
\def\ome{\omega}
\def\p{\pi}

\def\r{\rho}
\def\s{\sigma}
\def\t{\tau}

\def\D{\Delta}
\def\F{\Phi}
\def\G{\Gamma}
\def\J{\Psi}

\def\Ome{\Omega}

\def\S{\Sigma}

\def\cf{{\cal F}}

\def\cl{{\cal L}}

\def\cn{{\cal N}}
\def\co{{\cal O}}

\def\cs{{\cal S}}

\def\inbar{\vrule height1.5ex width.4pt depth0pt}
\def\rlx{\relax\leavevmode}
\def\I{\leavevmode\hbox{\small1\kern-3.8pt\normalsize1}}
\def\openone{\leavevmode\hbox{\small1\kern-3.3pt\normalsize1}}
\def\Ione{\rlx{\rm 1\kern-2.7pt l}}
\def\Ik{\rlx{\rm I\kern-.18em k}}
\def\IC{\rlx\leavevmode
             \ifmmode\mathchoice
                    {\hbox{\kern.33em\inbar\kern-.3em{\rm C}}}
                    {\hbox{\kern.33em\inbar\kern-.3em{\rm C}}}
                    {\hbox{\kern.28em\sinbar\kern-.25em{\rm C}}}
                    {\hbox{\kern.25em\ssinbar\kern-.22em{\rm C}}}
             \else{\hbox{\kern.3em\inbar\kern-.3em{\rm C}}}\fi}
\def\IP{\rlx{\rm I\kern-.18em P}}
\def\IR{\rlx{\rm I\kern-.18em R}}
\def\IN{\rlx{\rm I\kern-.20em N}}

\def\llsymbol#1{\@llsymbol{\@nameuse{c@#1}}}
\def\@llsymbol#1{\ifcase#1\or {}\or {'}\or {''}\or {'''}\or
   {''''}\or {'''''}\or  \else\@ctrerr\fi\relax}
\newcounter{contador}

\newcommand{\ol}\overline
\newcommand{\ti}\tilde
\newcommand{\wt}\widetilde
\newcommand{\wh}\widehat
\newcommand{\bv}\breve
\newcommand{\dg}\dagger

\newcommand{\C}{^{\mbox{\scriptsize c}}}

\renewcommand{\L}{_{\mbox{\scriptsize L}}}
\newcommand{\R}{_{\mbox{\scriptsize R}}}

\newcommand{\be}{\begin{equation}}
\newcommand{\ee}{\end{equation}}
\newcommand{\bl}{\begin{eqnarray}&}
\newcommand{\el}{&\end{eqnarray}}
\newcommand{\bq}{\begin{eqnarray}}
\renewcommand{\eq}{\end{eqnarray}}

\newcommand{\ov}{\overline}

\renewcommand{\pa}{\partial}

\renewcommand{\la}{\langle}
\newcommand{\ra}{\rangle}
\def\sl#1{\rlap{\hbox{$\mskip 1 mu /$}}#1}

\begin{document}
{\hfill\parbox{45mm}{{ 
hep-th/0012067\\
UFES-DF-OP2000/4\\
CBPF-NF-076/00\\
ICEN-PS-00/7
}} \vspace{3mm}

\begin{center}
{\LARGE {\bf Finiteness of PST self-dual models}} 

\vspace{7mm} 

{\large Oswaldo M. Del Cima$^{{\rm (a) }}$,
 Olivier Piguet$^{{\rm (b)1}}$ and Marcelo S. Sarandy
$^{{\rm (c)}}$\footnote{Supported by the Conselho Nacional 
de Desenvolvimento Cient\'{\i}fico e  
Tecnol\'{o}gico CNPq -- Brazil.}}   
\vspace{4mm}

$^{{\rm (a)}}$ {\it Universidade Cat\'olica de Petr\'opolis (UCP), \\ 
Grupo de F\'\i sica Te\'orica, \\Rua Bar\~ao do Amazonas 124 - 25685-070 - Petr\'opolis - RJ - Brazil.} 

\vspace{2mm}

$^{{\rm (b)}}$ {\it Universidade Federal do Esp\'\i rito Santo (UFES), \\
Departamento de F\'\i sica, \\Campus Universit\'ario de Goiabeiras - 
29060-900 - Vit\'oria - ES - Brazil.}

\vspace{2mm}

$^{{\rm (c)}}$ {\it Centro Brasileiro de Pesquisas F\'\i sicas (CBPF), \\
Coordena\c c\~ao de Teoria de Campos e Part\ii culas (CCP),\\
Rua Dr. Xavier Sigaud 150 - 
22290-180 - Rio de Janeiro - RJ - Brazil.}

\vspace{4mm} 

{\tt{E-mails:delcima@gft.ucp.br,
piguet@cce.ufes.br, sarandy@cbpf.br}} 
\end{center}

\begin{abstract}
The Pasti-Sorokin-Tonin model for describing chiral forms is 
considered at the quantum level. We study 
the ultraviolet and infrared behaviour of the model in two, four and six 
dimensions in the framework of algebraic renormalization. The absence of 
anomalies, as well as the finiteness, up to non-physical 
renormalizations, are shown in all dimensions analyzed.
\end{abstract}


\section{Introduction}

Antisymmetric tensor fields with self-dual field strengths, 
also called chiral
forms or chiral bosons, appear in a wide 
context in theoretical physics, mainly related to 
superstring theories~\cite{Gross}, M-theory 
five-branes~\cite{Witten}, chiral 
supergravity~\cite{Howe} and also in connection 
with fractional quantum Hall 
effect~\cite{Wen} and statistical systems involving 
the Coulomb Gas~\cite{West}.

Various formulations, non-manifestly~\cite{Floreanini}-\cite{Schwarz} 
and manifestly Lorentz covariant~\cite{Siegel}-\cite{Sorokin} 
have been proposed to describe 
chiral bosons. Among the manifestly Lorentz 
covariant versions, it can be pointed out the 
Siegel model~\cite{Siegel}, which imposes the 
square of the self-duality condition. It is   
affected by a gauge anomaly, what is extensively 
discussed in~\cite{Imbimbo}. In addition, 
one can still mention the two-dimensional 
covariant model proposed by 
McClaim-Wu-Yu~\cite{McClain} and Wotzasek~\cite{Wotzasek}, 
which introduces an 
infinite tower of auxiliary fields. This 
formulation has been generalized for 
higher dimensions in~\cite{Restuccia}.

Recently, a remarkable Lorentz 
covariant approach to describe chiral bosons, 
proposed by Pasti, Sorokin and Tonin (PST)~\cite{Pasti,Sorokin}, 
has been a subject of intensive 
research~\cite{Lechner1,Bandos,Lechner2}. 
It introduces, in contrast to~\cite{McClain,Wotzasek,Restuccia}, 
only one scalar auxiliary field, 
but in a non-polynomial way. The analysis is performed in 
Minkowski space-times with even dimension\footnote{See also the 
discussion about self-dual models 
defined in odd and even 
dimensional space-times by the mechanism 
of dual projection~\cite{Wotzasek2}.}, as usual 
in self-dual models. However, 
if the dimension is $D=4p$, with p integer, 
it is not possible to define a non-vanishing 
$2p$-form self-dual field-strength. So, in this case,   
a modified version of this approach has been defined. 
In particular, the PST model in four 
dimensions has been discussed~\cite{Pasti}, which 
describes a manifestly 
Lorentz and duality symmetric Maxwell electromagnetism. 
It is based on the duality symmetric 
-- but non-manifestly Lorentz invariant -- Schwarz-Sen 
model~\cite{Schwarz} and introduces, 
besides a non-polynomial dependence on a scalar auxiliary field, 
the presence of an auxiliary gauge potential. Then the
self-duality property is replaced by a duality relation between 
the two field-strengths of the theory.

The PST mechanism has been used, among 
other things, to construct a covariant 
effective action 
for the M-theory five-brane~\cite{Bandos}, 
covariant actions for chiral supersymmetric bosons~\cite{Lechner2} 
and to rederive the gravitational 
anomaly for the chiral bosons~\cite{Lechner1}.

Our purpose in this work is to investigate the 
quantum behaviour of the PST
models in the framework of the algebraic 
renormalization~\cite{Piguet}, analyzing in detail the
issue of the anomalies and the stability 
of the classical action under radiative
corrections. The absence of gauge anomaly  
has already been pointed out in two dimensions~\cite{Sorokin,Lechner1}. 
We give an algebraic proof of this result, 
including also the cases of four and six dimensions. 
Moreover we show the finiteness of the PST models, up 
to non-physical renormalizations, in all dimensions analyzed.   

The paper is organized as follows. 
In section 2 we describe the PST model in D=2 at the classical 
and quantum levels, discussing its infrared 
and ultraviolet properties, showing the absence 
of gauge anomaly as well as the finiteness. 
In section 3, we extend these results 
to the PST formulation in D=6. In Section 4 
we discuss the modified version of the PST approach 
in D=4, for which we show that the same results hold. 
In section 5 we summarize
our main results, presenting our conclusions.

\section{The PST formulation in D=2}

\subsection{The classical aspects}
The classical PST action in D=2 is defined as~\cite{Pasti}
\bq
\S_{\rm{inv}}\!\!\!&=&\!\!\! \int d^2x \,
\frac{1}{2}\left(-F^\m F_\m + \frac{1}{(\pa a)^2}\,
(\pa^{\m}a \,\cf_{\m})^{2}\right),
\label{invD2}
\eq
where $F_\m \equiv \pa_{\m}\f$ is the field strength 
of the scalar field $\f$, 
$\cf_\m \equiv (\h_{\m\n} - \e_{\m\n})\pa^{\n}\f=
-\e_{\m\n}\cf^{\n}$ and $a$ is a scalar 
field\footnote{We will work in Minkowski space-time with metric 
${\rm{diag}}(1,-1,...,-1)$. The Levi-Civita tensor 
$\e_{\m_{1}...\m_{D}}$ is defined by 
$\e^{01...D-1}=1=(-1)^{D-1}\e_{01...D-1}$.}.

The action (\ref{invD2}) is invariant under the following transformations :
\bq
&&\d_{\a}\f = \a(x) \ B,\,\,\,\d_{\a}a = \a(x)\ ,\quad
{\rm with}\  B \equiv
\frac{\cf_{\m}\pa^{\m}a} {(\pa a)^2}\,; \label{trans1D2} \\
&&\d_{\e}\f = f(a),\,\,\,\d_{\e}a = 0\ ,\quad 
{\rm{with}}\ f(a)\ {\rm{an}}\,
{\rm{arbitrary}}\,{\rm{function}}\,{\rm{of}}\,a\,. \label{trans2D2}
\eq
Actually the transformation (\ref{trans2D2}) corresponds to an 
infinite set of global symmetries. Using this invariance and the 
equation of motion for $\f(x)$ we can get the self-duality 
condition $\cf_{\m}=0$~\cite{Maznytsia}, which implies that $\f(x)$ is a 
chiral scalar.  By means of
the local symmetry (\ref{trans1D2}) we can choose $\pa_{\m}a=\d^{0}_{\m}$, 
getting the
non-manifestly Lorentz covariant Floreanini-Jackiw model~\cite{Floreanini}. 
However, as our result will show, the symmetry (\ref{trans1D2}), 
together just with the usual power-counting bounds for the dimension, 
is sufficient 
to fix the theory (up to field redefinitions) in the quantum regime 
as well as in the classical approximation. Hence we shall not need to 
include the invariance under (\ref{trans2D2}) among our basic 
requirements. The same observation holds for the higher dimensional 
models studied in the next sections.

In order to quantize the theory we must first determine the vacuum 
field configuration.
The auxiliary field $a$ enters in the action in a non-polynomial way and, 
to avoid a 
singularity, the condition  $\pa^{\m}a\pa_{\m}a \neq 0$ has to be imposed. 
So $a(x)$ is split as
\bq
a(x) = \bar{a}(x) + a^\prime(x),
\label{splitD2}
\eq
where $\bar{a}(x)$ is the vacuum expectation value of $a(x)$ 
and $a^\prime(x)$ is its quantum 
fluctuation.

We define the BRST transformations corresponding to the gauge symmetry 
(\ref{trans1D2}) as
\bq
&&s\f = c B ~~\nonumber\\
&&sa^\prime = c~~\nonumber\\
&&sc=0~~.
\label{BRSTD2}
\eq
where $c$ is the Faddeev-Popov ghost.

Moreover we have
\bq
s \S_{\rm{inv}} = 0 \ \ \ ,\ \ \ s^2 = 0.
\label{QinvD2}
\eq
In order to implement the gauge-fixing of the local 
symmetry (\ref{trans1D2}) 
let us introduce a Lagrange multiplier field $\p$ and an  
antighost field $\bar{c}$.
The gauge-fixing action is given by
\bq
\S_{\rm{gf}}\!\!\!&=&\!\!\! s \int{d^2x} \ \bar{c}a^\prime = 
\int{d^2x} \left(\p a^\prime 
- \bar{c}c \right),
\label{gfD2}
\eq
with
\bq
&&s\bar{c}=\p \ \ \ ,\ \ \ s\p=0~~.
\label{BRSTpcD2}
\eq
 In view of expressing the BRST invariance in a functional way by a 
Slavnov-Taylor identity we 
add to the action a term coupling an external 
field  $\f^*$ (``antifield''~\cite{Batalin}) to the BRST  
variation of $\f$,  
which is non-linear and hence subjected to possible renormalization:
\bq
\S_{\rm{ext}}\!\!\!&=&\!\!\! \int{d^2x} \ \f^*s\f = \int{d^2x}\ \f^{*}cB ,
\label{extD2}
\eq
An important point to be noticed concerns the infrared (IR) behaviour 
of the theory, which can be 
analyzed from the free propagators, listed below in 
momentum-space\footnote{Using the 
abbreviation 
notation $\la...\ra \equiv \la0|T...|0\ra_{\rm free}$, 
where $T$ is the time ordering operator.}
 (considering $\pa \bar{a}$ as a constant):
\bq
&&\la\f\f\ra= \left(-p^2 + 
\frac{\pa^{\m}\bar{a}\pa^{\n}\bar{a}}{(\pa\bar{a})^2}(p_{\m}p_{\n}-
2\e_{\n\g}p_{\m}p^{\g}+\e_{\m\l}\e_{\n\g}p^{\l}p^{\g})\right)^{-1}~~,\\
&&\la\bar{c}c\ra=1~~,\\
&&\la a^\prime \p\ \ra=1~~.
\label{propD2}
\eq
As we can see the propagator $\la\f\f\ra$ is not integrable at 
small momenta. This is a 
typical behaviour of the massless scalar fields in 
two space-time dimensions.
So we must regularize the ill defined propagator at long distances and, in
order to do so, we introduce a regulator  mass~\cite{Blasi} such that
\bq
\la\f\f\ra=\left(-p^2 + 
\frac{\pa^{\m}\bar{a}\pa^{\n}\bar{a}}{(\pa\bar{a})^2}(p_{\m}p_{\n}-
2\e_{\n\g}p_{\m}p^{\g}+\e_{\m\l}\e_{\n\g}p^{\l}p^{\g}) + 
m_{\f}^2\right)^{-1}~~,
\label{propD2Reg}
\eq
where $m_{\f}^2$ is a regulator mass, giving a well defined 
behaviour for the 
propagator in the IR limit.

Such a mass may be introduced, formally keeping BRST invariance, through 
the mass action
\bq
\S_{\rm{m}}\!\!\!&=&\!\!\! s \int{d^2x}\ \frac{1}{2} \t_{2}\f^2 =
 \int{d^2x} \left(
\frac{1}{2}(\t_{1}+m_{\f}^{2})\f^2 - \t_{2}\f cB\right),
\label{massD2}
\eq
where $\t_{1}$ and $\t_{2}$ are new external fields with the 
following BRST
transformations :
\bq
&&s\t_{2} = \t_{1} + m_{\f}^2\ \ \ , \ \ \ s\t_{1} = 0.
\label{tsourD2}
\eq
So, the total classical action for the PST model in D=2,
\bq
\S\!\!\!&=&\!\!\!\S_{\rm{inv}} + \S_{\rm{gf}} +\S_{\rm{ext}}
+\S_{\rm{m}}\ ,
\label{actionD2}
\eq
is invariant under BRST transformations, what is expressed in 
a functional way 
by the Slavnov-Taylor identity
\bq
{\cal S}(\S)\!\!\!&=&\!\!\! \int {d^2x}\left(\frac{\d\S}{\d\f^{*}}
\frac{\d\S}{\d\f}+c{\frac{\d\S}{\d a^\prime}}+\p{\frac{\d\S}{\d\bar{c}}}+
(\t_{1}+m_{\f}^{2}){\frac{\d\S}{\d\t_{2}}}\right)=0~~.
\label{slavnovD2}
\eq
  From (\ref{slavnovD2}) we define the linearized Slavnov-Taylor operator 
as below
\bq
\cs_{\S}\!\!\!&=&\!\!\!\int{d^2x} 
\left(\frac{\d\S}{\d\f^{*}}\frac{\d}{\d\f}+
\frac{\d\S}{\d\f}\frac{\d}{\d\f^*}+c{\frac{\d}{\d a^\prime}}+
\p{\frac{\d}{\d\bar{c}}}+(\t_{1}+m_{\f}^{2}){\frac{\d}{\d\t_{2}}}\right)~~.
\label{linSlavD2}
\eq
For any functional $\g$ it can be shown to hold 
${\cal S}_\g{\cal S}(\g) = 0$. 
Moreover taking into account the Slavnov-Taylor 
identity (\ref{slavnovD2}), the 
linearized Slavnov-Taylor operator turns out to be nilpotent :
\bq
\cs_{\S}\cs_{\S}\!\!\!&=&\!\!\!0.
\eq
The ultraviolet (UV) and infrared (IR) dimensions, as well as 
the Faddeev-Popov charge 
of all fields are displayed in Table 1.
\begin{table}[hbt]
\centering
\begin{tabular}{|c||c|c|c|c|c|c|c|c|c|}
\hline
\phantom{$\dfrac{x}{x}$}
& $\f$ & $a^{\prime}$ & $c$ & $\p$ & $\bar{c}$ &  
$\f^*$ & $\t_1$ & $\t_2$ & $s$\\ \hline\hline
$d$ & $0$ & $-1$ & $-1$ & $3$ & $3$ & $2$ & 
$2$ & $2$ & $0$ \\ 
\hline
$r$ & $1$ & $-1$ & $-1$ & $3$ & $3$ & $2$ & 
$2$ & $2$ & $0$ \\ 
\hline
$\F\Pi$ & 0 & 0 & 1 & $0$ & $-1$ & $-1$ & 0 & $-1$ & 1\\ \hline
\end{tabular}
\caption[t1]{UV and IR dimensions $d$ and $r$, ghost number $\F\Pi$.}
\label{table1}
\end{table}

Before discussing the quantum behaviour of the model let us note 
important constraints, besides the Slavnov-Taylor
identity, obeyed by the total classical action (\ref{actionD2}). 
The gauge condition and 
the ghost equation read
\bq
&&{\frac{\d\S}{\d\p}}=a^\prime~~ \ \ \ ,\ \ \
 {\frac{\d\S}{\d\bar{c}}}=-c~~ .
\label{ghostgaugeD2}
\eq 
The dependence of the total classical action (\ref{actionD2}) 
on $a^\prime$ and $\bar{a}$ is
essentially through the combination $a^\prime + \bar{a}$, what is 
expressed by the following constraint :
\bq
&&\left({\frac{\d}{\d a^\prime}}-{\frac{\d}{\d\bar{a}}}\right)\S = \p .
\label{constraintD2}
\eq
where the r.h.s. is a classical insertion, {\it i.e.}, a linear 
term in the
quantum fields which thus will not get renormalized.

Moreover, important constraints are given by the integrated equations of 
motion of $a^\prime$  and $\f$, which 
imply that the total classical action 
depends  essentially on the derivatives of these fields. 
These equations are
given by  
\bq
&&\int d^2x \
\frac{\d\S}{\d a^{\prime}} = \int d^2x \ \p , \label{eqmotaD2} \\ 
&&\int d^2x \
\left(\frac{\d}{\d\f}+\t_{2}\frac{\d}{\d\f^*}\right)\S =  \int d^2x \
\left(\t_{1}+m_{\f}^{2}\right)\f . 
\label{eqmotfD2} 
\eq
where the r.h.s. of these equations are also classical insertions.

\subsection{The quantum aspects}
This subsection is devoted to study the possibility of 
implementing the Slavnov-Taylor 
identity at the quantum level and the stability of 
the classical action under quantum corrections. 
This amounts to study all possible anomalies and invariant 
counterterms in the total action.

First, it is important to mention that there is no problem in the quantum
extension of constraints of the 
type (\ref{ghostgaugeD2} - \ref{eqmotfD2})~\cite{Piguet}.

In order to show that the Slavnov-Taylor identity can be 
implemented at the quantum level
we will start by applying the quantum action principle 
(QAP)~\cite{Piguet,Lowenstein}, 
which assures that any possible breaking $\D^{(1)}$ of the 
Slavnov-Taylor identity takes the following form :
\bq {\cal S}(\G) = \D^{(1)}\cdot\G = \D^{(1)}+\co(\hbar\D^{(1)})~~.
\label{breakingD2}
\eq
where $\G$ is the vertex functional and $\D^{(1)}$ is a local, 
integrated, Lorentz invariant 
polynomial of UV dimension $\leq 2$, IR dimension $\geq 2$ 
and ghost number $1$.
 
The identity $\cs_{\G}{\cal S}(\G) = 0$ and the fact that 
$\cs_{\G}=\cs_{\S} + \co(\hbar)$, 
lead to the Wess-Zumino consistency condition for the breaking $\D^{(1)}$ :
\bq
\cs_{\S} \D^{(1)} = 0 ,
\label{WZD2}
\eq 
which constitutes a cohomology problem. The proof of 
the absence of anomaly consists in
showing that $\D^{(1)}$ is in the trivial sector of the cohomology, 
{\it i.e.}, $\D^{(1)}=
\cs_{\S}\hat{\D}^{(0)}$, where $\hat{\D}^{(0)}$ is a local 
polynomial in the fields, 
with ghost number $0$, called noninvariant counterterm.

In order to characterize all possible invariant counterterms $\D^{(0)}$, 
considered as 
perturbations of the total classical action, we are led to solve again 
a cohomology equation :
\bq
\cs_{\S} \D^{(0)} = 0 ,
\label{StabD2}
\eq   
where $\D^{(0)}$ is a local, integrated, Lorentz invariant polynomial 
of UV dimension 
$\leq 2$, IR dimension $\geq 2$ and ghost number $0$.

Anomalies and counterterms $\D^{(G)}$ (where $G=0,1$ 
denotes the ghost number) have also to 
obey the conditions :  
\bq
&&{\frac{\d\D^{(G)}}{\d\p}} = 0~~\ \ ,\ \ 
{\frac{\d\D^{(G)}}{\d\bar{c}}} = 0~~, \nonumber \\ 
&&\left({\frac{\d}{\d a^\prime}} -
 {\frac{\d}{\d\bar{a}}}\right)\D^{(G)}= 0, \nonumber \\
&&\int d^2x \ \frac{\d\D^{(G)}}{\d a^{\prime}} = 0 , \nonumber \\
&&\int d^2x \ \left(\frac{\d}{\d\f}+
\t_{2}\frac{\d}{\d\f^*}\right)\D^{(G)} = 0. 
\label{condCTD2}
\eq
which follow from requiring the fulfillment of the constraints 
(\ref{ghostgaugeD2} - \ref{eqmotfD2}) 
to all orders.

In order to determine the cohomology of $\cs_{\S}$ one introduces the 
counting operator 
\bq
\cn = \int d^2x \sum_{i} \F_{i}\frac{\d}{\d\F_{i}},
\label{countopD2}
\eq
with $\F_{i}$ running over all fields. So one can decompose $\cs_{\S}$ as
\bq
\cs_{\S} = \cs_{\S}^0 + \cs_{\S}^1 + ...
\label{decslavD2}
\eq
where $\cs_{\S}^n$ is the contribution of power $n$ in the fields for 
$\cs_{\S}$, defined by means of $[\cn,\cs_{\S}^n]=n\ \cs_{\S}^n$.
The $\cs_{\S}^0$-transformations read 
\be\ba{ll}
\cs_{\S}^0 a^\prime = c\ ,
   & \cs_{\S}^0 c = 0\ , \nonumber \\
\cs_{\S}^0 \f = 0\ ,
   & \cs_{\S}^0 \f^* = \left. 
\dfud{\S}{\f}\right|_{\rm linear\,\,approx.}\  , \nonumber \\
\cs_{\S}^0 \bar{c} = \p\ ,
   & \cs_{\S}^0 \p = 0\ , \nonumber \\
\cs_{\S}^0 \t_{2} = \t_{1} + m_{\f}^2\ , \quad\quad
   &\cs_{\S}^0 \t_{1} = 0\ . 
\label{BRST0D2}
\ea\ee
According to a general theorem~\cite{Piguet,Dixon,Becchi}, 
the cohomology of the complete operator $\cs_{\S}$ 
is isomorphic to a subspace of the cohomology of the operator 
$\cs_{\S}^0$. Thus we 
will first focus on the operator $\cs_{\S}^0$. Studying the most 
general dependence of
the cohomology on the antifield $\f^*$ we write down
\bq
\D^{(G)} = \int d^2x \, \f^* Z^{(G+1)} + 
 \,\rm{terms}\,\rm{independent}\,\rm{of}\,\rm{the} 
\,\rm{antifield}\,.
\label{CounterAFD2}
\eq
where $Z^{(G+1)}$ is an arbitrary local, 
non-integrated, Lorentz invariant polynomial of UV
dimension $\leq 0$, IR dimension $\geq 0$ and ghost number
 $G+1$ depending 
on all fields 
except $\f^*$. Hence from (\ref{WZD2}) and 
(\ref{StabD2}), at lowest order in $n$, we have: 
\bq
\cs_{\S}^0 \D^{(G)}&=& 0 \nonumber \\
&=& - \int d^2x \, \f^* \cs_{\S}^0 Z^{(G+1)} + 
\,\rm{terms}\,\rm{independent}\,\rm{of}\,\rm{the}\,\rm{antifield}\,,
\label{CohomAFD2}
\eq
  From (\ref{CohomAFD2}) we get a new (local) cohomology problem: 
\bq
\cs_{\S}^0 \ Z^{(G+1)}(x) =  0\ .
\label{localcohD2}
\eq
We observe in (\ref{BRST0D2}) that various fields appear 
as BRST doublets 
$(\vf_1,\,\vf_2=\cs_{\S}^0\vf_1)$, namely 
$(\bar{c},\,\p)$, $(\t_{2},\,\t_{1}+m_{\f}^{2})$, $(a',\,c)$, as well 
all their derivatives. As it 
is well known ~\cite{Piguet,Dixon}, such doublets belong to the 
trivial sector of the cohomology.

Since $c$ and its derivatives in particular are in doublets, 
there are no fields with positive ghost 
number in the non-trivial sector of the cohomology, which thus is 
empty for $G=0,1$. 

Let us now analyze the terms independent of $\f^*$. This means to study
\bq
\cs_{\S}^0 \int d^2x \ Q^{(G)} = 0
\label{CohomFD2}
\eq
where $Q^{(G)}$ has UV dimension $\leq 2$, IR dimension $\geq 2$ 
and ghost number $G=1$ for 
anomalies and $G=0$ for the invariant counterterms.

  From (\ref{CohomFD2}) and the algebraic Poincar\'e 
lemma~\cite{Henneaux2} one deduces the 
set of descent equations
\bq
&&\cs_{\S}^0 Q^{(G)} = \pa^{\m} Q_{\m}^{(G+1)}~~,\nonumber \\
&&\cs_{\S}^0 Q_{\m}^{(G+1)} \ = \pa^{\n} Q_{[\m\n]}^{(G+2)}~~,\nonumber \\
&&\cs_{\S}^0 Q_{[\m\n]}^{(G+2)} = 0
\label{decohomD2}
\eq
There is no non-trivial solution with ghost number $G=1$, what means 
that the 
cohomology of $\cs_{\S}$ with $G=1$ is empty and the model is anomaly-free.

Concerning the possible invariant counterterms ($G=0$) we get that the 
most general non-trivial solution
for the top level of the descent equations is given by 
\bq
&&Q^{(0)} = \Ome(\f,\bar{a}),
\label{Cohomols0D2}
\eq
where $\Ome(\f,\bar{a})$ is an arbitrary polynomial of $\f$ -- 
with coefficients as arbitrary functions of $\bar{a}$ -- with UV dimension 
$\leq2$, IR dimension $\geq2$ and ghost number $0$.

In order to determine the cohomology of the full operator 
$\cs_{\S}$, we must first complete 
the solution (\ref{Cohomols0D2}) to an invariant of $\cs_{\S}$,
imposing the constraints 
(\ref{condCTD2}). So we get that the most general non-trivial 
counterterm can be at most  
the invariant action (\ref{invD2}) :
\bq
&&\int d^2x \ Q^{(0)} = \S_{\rm{inv}}.
\label{CohomolsD2}
\eq
However $\S_{\rm{inv}}$ is trivial :
\bq
&&\S_{\rm{inv}} = \frac{1}{2} \ \cs_{\S} \int d^2x \ \f \f^* - 
\S_{\rm{m}}\ .
\label{StrivD2}
\eq
since the mass term $\S_{\rm{m}}$ is trivial, according to \equ{massD2}.
Therefore, the cohomology of $\cs_{\S}$ in the sector $G=0$ is empty and 
the PST model in $D=2$ is finite, up to possible non-physical 
renormalizations of the field amplitudes which 
correspond to the trivial solution of the cohomology equation 
(\ref{StabD2}).

It is important to mention that it remains the question of the 
massless limit $m_{\f}\to 0$, 
since $m_{\f}$ is not a physical mass and has been introduced just as 
an infrared cut-off. This kind of 
discussion can be found in~\cite{Becchi} in the context 
of the two dimensional 
non-linear sigma model and in~\cite{Blasi} in connection with the BF model.

\section{Finiteness of the D=6 model}

\subsection{The classical model}
The extension of the PST action to dimension 
D=6 is given by~\cite{Pasti,Sorokin}
\bq
\S_{\rm{inv}}\!\!\!&=&\!\!\! \int d^6x \ 
\left(-\frac{1}{6}F^{\m\n\r} F_{\m\n\r} 
+ \frac{1}{2(\pa a)^2}\ \pa^{\m}a \ \cf_{\m\n\r} 
\cf^{\n\r\l} \pa_{\l}a \right),
\label{invD6}
\eq
where $F_{\m\n\r} \equiv \pa_{\m}A_{\n\r} + \pa_{\r}A_{\m\n} + 
\pa_{\n}A_{\r\m}$ is 
the field strength of the tensorial field $A_{\m\n}$, 
\bq
\cf_{\m\n\r} \equiv F_{\m\n\r} - 
\frac{1}{3!} \e_{\m\n\r\s\e\d}F^{\s\e\d} = 
-\frac{1}{3!} \e_{\m\n\r\s\e\d}\cf^{\s\e\d}\ ,
\eq
and $a$ is a scalar auxiliary field.

The action (\ref{invD6}) is invariant under the following 
set of four transformations:
\be\ba{lll}
\d_{I}A_{\m\n} = 2\,\pa_{[\m} \a_{\n]} 
   = \pa_{\m} \a_{\n} - \pa_{\n} \a_{\m}\ ,\quad& \d_{I}a= 0 \ ,&
\label{invID6} \es
\d_{II}A_{\m\n} = 2\,\f_{[\m} \pa_{\n]}a  
    = \f_{\m} \pa_{\n}a - \f_{\n} \pa_{\m}a \ ,\quad& \d_{II}a= 0 \ ,&
\label{invIID6} 
\es
\d_{III}A_{\m\n} = \b B_{\m\n} \ ,\quad& \d_{III}a= \b\ ,
    \quad& {\rm{with}}\ B_{\m\n}
          \equiv \dfrac{\cf_{\m\n\r}\pa^{\r}a}{(\pa a)^2}\ , 
\label{invIIID6} 
\es
\d_{IV}A_{\m\n} = f_{\m\n}(a)\ ,\quad& \d_{IV}a= 0 \ ,
     \quad& {\rm{with}}\,\,f_{\m\n}(a)\,{\rm{an}}\,
                                     {\rm{arbitrary}}\\
& & \,{\rm{function}}\,{\rm{of}}\,a\ ,
\label{invIVD6}
\ea\ee
where $\a_{\m}$, $\f_{\m}$ and $\b$ are the infinitesimal 
parameters of the transformations.

As in D=2 we must require $\pa^{\m}a\pa_{\m}a \neq 0$ and so 
split $a(x)$ in $\bar a(x)$ 
and $a^\prime(x)$ according to (\ref{splitD2}).

The BRST transformations, arising from 
the local symmetries of the set (\ref{invIVD6}) 
are given by
\bq
&&sA_{\m\n} = 2\,\pa_{[\m} \a_{\n]} + 2\,\f_{[\m} \pa_{\n]}a + 
\b B_{\m\n},~~\nonumber\\
&&sa^\prime = \b,~~\nonumber\\
&&s\b=0~~,\nonumber\\
&&s\a_{\m} = \pa_{\m}\a + \pa_{\m}a \ome - \f_{\m}\b~~,\nonumber\\
&&s\f_{\m} = -\pa_{\m}\ome - \pa_{\m}a\f + 2\,\pa_{[\m} \f_{\n]}
\frac{\pa^{\n}a \b}{(\pa a)^2} - B_{\m\n}
\frac{\b\pa^{\n}\b}{(\pa a)^2}~~,\nonumber\\
&&s\a = -\ome\b~~,\nonumber\\
&&s\ome = -\f\b~~,\nonumber\\
&&s\f = -B_{\m\n}\frac{\b \pa^{\m}\b \pa^{\n}\b}{(\pa a)^4} + 
2\,\frac{\pa^{\m}a}{(\pa a)^4} 
\pa_{[\m} \f_{\n]} \b \pa^{\n}\b + 
\frac{\pa^{\m}a}{(\pa a)^2}\pa_{\m}\f\b.
\label{BRSTD6} 
\eq
where $\a_{\m}$, $\f_{\m}$ and $\b$ are now anticommuting 
ghosts with ghost number one. 
The fields $\a$, $\ome$ and $\f$ are commuting 
ghosts with ghost number two, which were introduced to fix 
the residual degrees of freedom 
coming from the reducible symmetries 
(the first two symmetries of the set (\ref{invIVD6})). The latter have 
indeed three zero modes: $\a_{\m}=\pa_{\m}X$; $\f_{\m}=\pa_{\m}a Y$; 
$\a_{\m}=\pa_{\m}a Z$, $\f_{\m}=-\pa_{\m}Z$. 

In addition we have
\bq
s \S_{\rm{inv}} = 0 \ \ ,\ \ s^2 = 0.
\label{sinvD6}
\eq
Following the Batalin-Vilkovisky prescription~\cite{Batalin} for 
theories with 
reducible symmetries we implement the gauge-fixing action
\bq
&&\S_{\rm{gf}} = s \int{d^6x}\left(\bar\a^{\m}\pa^{\n}A_{\m\n} + 
\bar\f^{\m}A_{\m\n}\pa^{\n}a + \bar{\f}^\prime \pa^{\m}a \f_{\m} +
\bar \a
\pa^{\m}\a_{\m} +  \right. \nonumber \\
&&\left. \hspace{2.3cm}+ \bar\f\pa^{\m}\f_{\m} + \bar\b a^\prime +
\bar\a^{\m}\pa_{\m}\bar c+\bar{\f}^{\m}\pa_{\m}a \bar{c}^\prime\right),
 \label{gfD6}
\eq
where we have introduced the antighosts $\bar\a^{\m}$, $\bar\f^{\m}$,
 $\bar\a$, $\bar\f^\prime$, $\bar\f$ and
$\bar\b$, the Lagrange multipliers $\p^{\m}$,
 $\r^{\m}$, $\p$, $\r^\prime$, $\r$ and $b$ and the extra pairs of
fields $(\bar c,\l)$ and  $(\bar{c}^\prime,\l^\prime)$ transforming as
\bq
&&s\,{\bar{C}} = \Pi \ , \quad s\,\Pi = 0~,
\label{BRSTpcD6}
\eq
with $\bar{C} = (\bar\a^{\m},\,\bar\f^{\m},\,\bar{\a},\,
\bar\f^{\prime},\,\bar\f,\,\bar\b
,\,\bar{c},\,\bar{c}^\prime)$ and $\Pi =
 (\p^{\m},\,\r^{\m},\,\p,\,\r^{\prime},\,
\r,\,b,\,\l,\,\l^\prime)$.

Differently from the case D=2 there is no problem in the
 infrared limit because all the
propagators turn out to be integrable at small momenta in 
D=6. The UV dimensions 
and ghost numbers of all fields introduced so far are shown 
in Table 2 and Table 3.

\begin{table}[hbt]
\centering
\begin{tabular}{|c||c|c|c|c|c|c|c|c|c|}
\hline
\phantom{$\dfrac{x}{x}$}
& $A_{\m\n}$ & $a^{\prime}$ & $\a_\m$ & $\f_\m$ & $\b$ & $\a$ &
          $\ome$ & $\f$ & $s$\\ \hline\hline
$d$ & $2$ & $-1$ & $1$ & $2$ & $-1$ & $0$ & $1$ & 
$2$ & $0$ \\ 
\hline
$\F\Pi$ & $0$ & $0$ & $1$ & $1$ & $1$ & $2$ & $2$ & $2$ & $1$\\ \hline
\end{tabular}
\caption[t1]{dimension $d$ and ghost number $\F\Pi$.}
\label{table2}
\end{table}

\begin{table}[hbt]
\centering
\begin{tabular}{|c||c|c|c|c|c|c|c|c|c|c|c|c|c|c|c|c|}
\hline
\phantom{$\dfrac{x}{x}$}
& $\bar\a^\m$ & $\p^\m$ & $\bar\f^\m$ & $\r^\m$ & $\bar\a$ & $\p$ &
${\bar\f}^\prime$ &  $\r^\prime$ & 
$\bar\f$ & $\r$ & $\bar\b$ & $b$ & $\bar{c}$ & $\l$ & 
$\bar{c}^\prime$ & $\l^\prime$ \\ \hline\hline
$d$ & $3$ & $3$ & $4$ & $4$ & $4$ & $4$ & $4$ & $4$ & $3$ & 
$3$ & $7$ & $7$ & $2$ & $2$ & $2$ & $2$\\ 
\hline
$\F\Pi$ & $-1$ & $0$ & $-1$ & $0$ & $-2$ & $-1$ & $-3$ & $-2$ &
 $-2$ & $-1$ & $-1$ & $0$ & $0$ & $1$ 
& $0$ & $1$\\ \hline
\end{tabular}
\caption[t1]{dimension $d$ and ghost number $\F\Pi$.}
\label{table3}
\end{table}

In order to get the Slavnov-Taylor identity we introduce the
 external field action
\bq
\S_{\rm{ext}}\!\!\!&=&\!\!\! \int{d^6x} 
\left( \frac{1}{2}A_{\m\n}^{*}sA^{\m\n} + \a^{*}_{\m}s\a^{\m} +
\f^{*}_{\m}s\f^{\m} + \a^{*}s\a + \ome^{*}s\ome + \f^{*}s\f \right),
\label{extD6}
\eq
where $A_{\m\n}^{*}$, $\a^{*}_{\m}$, $\f^{*}_{\m}$, $\a^{*}$,
 $\ome^{*}$ and $\f^{*}$ are external 
sources, with UV dimensions and ghost numbers given in Table 4.

\begin{table}[hbt]
\centering
\begin{tabular}{|c||c|c|c|c|c|c|}
\hline
\phantom{$\dfrac{x}{x}$}
& $A_{\m\n}^{*}$ & $\a_\m^{*}$ & $\f_\m^{*}$ & $\a^{*}$ & 
$\ome^{*}$ & $\f^{*}$ \\ \hline\hline
$d$ & $4$ & $5$ & $4$ & $6$ & $5$ & $4$ \\ 
\hline
$\F\Pi$ & $-1$ & $-2$ & $-2$ & $-3$ & $-3$ & $-3$ \\ \hline
\end{tabular}
\caption[t1]{dimension $d$ and ghost number $\F\Pi$.}
\label{table4}
\end{table}
So, the total action is
\bq
\S\!\!\!&=&\!\!\!\S_{\rm{inv}} + \S_{\rm{gf}} +\S_{\rm{ext}}
\label{actionD6}
\eq
and the Slavnov-Taylor identity is written down as
\bq
{\cal S}(\S)\!\!\!&=&\!\!\! \int {d^6x}
\left({\frac{1}{2}}{\frac{\d\S}{\d A_{\m\n}^{*}}
\frac{\d\S}{\d A^{\m\n}}}+{\frac{\d\S}{\d\a_{\m}^{*}}
\frac{\d\S}{\d\a^{\m}}}+
{\frac{\d\S}{\d\f_{\m}^{*}}\frac{\d\S}{\d\f^{\m}}}+
{\frac{\d\S}{\d\a^{*}}\frac{\d\S}{\d\a}}+
{\frac{\d\S}{\d\ome^{*}}\frac{\d\S}{\d\ome}}+
{\frac{\d\S}{\d\f^{*}}\frac{\d\S}{\d\f}}~+\right. \nonumber \\
&&+\left.
\b{\frac{\d\S}{\d a^\prime}}+\p^\m{\frac{\d\S}{\d\bar\a^{\m}}}+
\r^\m{\frac{\d\S}{\d\bar\f^{\m}}}+\p{\frac{\d\S}{\d\bar\a}}+
 \r^\prime{\frac{\d\S}{\d\bar\f^\prime}}+
\r{\frac{\d\S}{\d\bar\f}}+b{\frac{\d\S}{\d\bar\b}}+
\l{\frac{\d\S}{\d \bar c}}+\l^\prime{\frac{\d\S}{\d \bar{c}^\prime}}
 \right)=0.~~
\label{slavnovD6}
\eq
  From the Slavnov-Taylor identity we get the nilpotent linearized 
Slavnov-Taylor operator
\bq
{\cs_{\S}}\!\!\!&=&\!\!\! \int {d^6x}
\left(\frac{1}{2}\frac{\d\S}{\d A_{\m\n}^*}\frac{\d}{\d A^{\m\n}}+
\frac{1}{2}\frac{\d\S}{\d A^{\m\n}}\frac{\d}{\d A_{\m\n}^*}+
\frac{\d\S}{\d\a_{\m}^*}\frac{\d}{\d\a^\m}+
\frac{\d\S}{\d\a^{\m}}\frac{\d}{\d\a_{\m}^*}+
\frac{\d\S}{\d\f_{\m}^*}\frac{\d}{\d\f^\m}~+\right. \nonumber \\
&&+\left.\frac{\d\S}{\d\f^\m}\frac{\d}{\d\f_{\m}^{*}}+
\frac{\d\S}{\d\a^{*}}\frac{\d}{\d\a}+
\frac{\d\S}{\d\a}\frac{\d}{\d\a^{*}}+
\frac{\d\S}{\d\ome^{*}}\frac{\d}{\d\ome}+
\frac{\d\S}{\d\ome}\frac{\d}{\d\ome^*}+
\frac{\d\S}{\d\f^{*}}\frac{\d}{\d\f}+
\frac{\d\S}{\d\f}\frac{\d}{\d\f^{*}}~+\right. \nonumber \\
&&+\left.
\b\frac{\d}{\d a^\prime}+\p^\m\frac{\d}{\d\bar\a^{\m}}+
\r^\m\frac{\d}{\d\bar\f^{\m}}+
\p\frac{\d}{\d\bar\a}+\r^\prime{\frac{\d}{\d\bar\f^\prime}}+
\r\frac{\d}{\d\bar\f}+b\frac{\d}{\d\bar\b}+
\l\frac{\d}{\d\bar{c}}+\l^\prime{\frac{\d}{\d \bar{c}^\prime}}\right). ~~
\label{LinSlavD6}
\eq
The total classical action (\ref{actionD6}) obeys various 
constraints which can be extended 
to the quantum level. Let us establish 
some of them, which will play a special role. First, 
the integrated antighost equation for $\a^\m$ and the 
non-integrated antighost equation for $\a$ are
\be
\int d^6x \ \frac{\d\S}{\d\a^{\m}} = 0\ ,\qquad
\frac{\d\S}{\d\a} = \pa^\m\pa_\m \bar\a - \pa^\m\a^*_\m\ .
\label{aghostcondD6}
\ee
The integrated equation of motion for $a^\prime$, which expresses, 
up to a classical breaking, 
that the action 
depends on $a^\prime$ only through its derivatives, reads
\bq
&&\int d^6x \ \frac{\d\S}{\d a^{\prime}} = \int d^6x \ b.
\label{eqmotaD6}
\eq
Finally, there is a constraint between $a$ and $a^\prime$, 
which is also classically broken and 
expresses that the action depends essentially on the combination 
$\bar{a} + a^\prime$. It is 
given by
\bq
&&\left({\frac{\d}{\d a^\prime}}-{\frac{\d}{\d \bar{a}}}\right)\S = b.
\label{constraintD6}
\eq
\subsection{Absence of anomalies and finiteness}
As it was recalled in Section 2, the possible anomalies and 
non-trivial invariant counterterms 
are characterized by the following cohomology problem
\bq
\cs_{\S} \D^{(G)} =0,
\label{CohomolD6}
\eq
where $\D^{(G)}$ is an integrated, Lorentz invariant,
 local polynomial of UV dimension $6$ 
and ghost number $G$, with the candidates for anomalies 
in the sector $G=1$ and the possible 
invariant counterterms in the sector $G=0$.

According to~\cite{Piguet} the constraints 
(\ref{aghostcondD6} - \ref{constraintD6}) can be 
implemented at the quantum level. 
It follows that $\D^{(G)}$ obeys the same 
constraints, but with their right-hand sides set to zero. Hence, in 
particular,  it will 
depend on $\a_{\m}$ through its derivatives only, and it will not 
depend on $\a$. 

In order to solve (\ref{CohomolD6}), we expand $\cs_{\S}$ according 
to a counting operator, 
like (\ref{countopD2}), acting on all fields. So, we can write down 
the following expansion 
for $\cs_{\S}$ :
\bq
\cs_{\S} = \ov\cs_{\S}^0 + \ov\cs_{\S}^1 + ...
\label{decslavD6}
\eq
However, instead of simply studying the 
cohomology of $\ov\cs_{\S}^0$ as we did 
in the case D=2, we will introduce two new 
filtrations, one of them corresponding to an 
expansion in powers of $\f$ and the other 
in terms of ($\f_{\m}$, $\ome$, $\a$). 
Denoting the corresponding $n$-th order
parameters as $\tilde{\cs_{\S}^n}$ and $\cs_{\S}^n$, 
respectively, we get
\bq
&&\ov\cs_{\S}^0 = \tilde{\cs_{\S}^0} + \tilde{\cs_{\S}^1} \ , \nonumber \\
&&\tilde{\cs_{\S}^0} = \cs_{\S}^0 + \cs_{\S}^1 \ ,
\label{newfilterD6}
\eq  
with the expansions stopping in order $1$ since $\ov\cs_{\S}^0$ and 
$\tilde{\cs_{\S}^0}$ are at most linear in $\f$ and  
($\f_{\m}$, $\ome$, $\a$), respectively.

 These new filtrations are very suitable, as it will be seen below, 
since they produce more doublet pairs, all of them being left
in the trivial sector of the cohomology.

Our aim now is to show that the cohomology of $\cs_{\S}^0$ is trivial, 
which implies 
that the cohomology of $\cs_{\S}$ is empty. 
The $\cs_{\S}^0$-transformations are
\be\ba{ll}
\cs_{\S}^0 A_{\m\n} = 2 \pa_{[\m} \a_{\n]} \ , 
  & \cs_{\S}^0 \a_\m = 0\ ,\nonumber \\
\cs_{\S}^0 a^\prime = \b \ ,        & \cs_{\S}^0 \b = 0\ , \nonumber \\
\cs_{\S}^0 \f = 0 \ ,     & \cs_{\S}^0 \a = 0\ , \nonumber \\
\cs_{\S}^0 \f_\m = -\pa_\m \ome\ ,\ &\cs_{\S}^0 \ome = 0\ , \nonumber \\
\cs_{\S}^0 \F^* = \left. \dfrac{\d\S}{\d\F} \right|_{\rm order\,\,0}\ ,
\label{B0transfD6}
\ea\ee
with  $\F^*=(A_{\m\n}^*$, $\f_{\m}^*$, $\a_{\m}^*$, $\a^*$,
 $\ome^*$, $\f^*)$ 
and $\F=(A_{\m\n}$, $\f_{\m}$, $\a_{\m}$, $\a$, $\ome$, $\f)$.

The transformations of the antighosts, Lagrange multipliers and 
pairs of extra fields 
$(\bar{c},\l)$ and $(\bar{c}^\prime,\l^\prime)$ 
remain identical to their BRST transformations 
as listed in (\ref{BRSTpcD6}).

Besides the doublets involving the antighosts, the 
Lagrange multipliers and the extra 
fields, there is a further doublet in (\ref{B0transfD6}), namely
$(a^\prime,\b)$, which eliminates the possibility of having fields 
with negative dimension in the cohomology of $\cs_{\S}^0$.
 Moreover, {\it in the space of non-integrated local polynomials}, 
$\f_{\m}$ and its symmetric derivatives compose 
doublets with derivatives of $\ome$.
  
Beginning with the cocycles involving the external fields, we can write
down the most general dependence on the latters as 
\bq
&&\D_{\rm{ext}}^{(G)}= \int d^6x \
 \left(\frac{1}{2}A_{\m\n}^{*}X^{(G+1)\m\n} 
+ \a_{\m}^{*} X^{(G+2)\m} + \f_{\m}^{*} 
Y^{(G+2)\m} + \a^{*} X^{(G+3)} + \ome^{*} Y^{(G+3)} +~\right.\nonumber \\ 
&&\left.\hspace{2.5cm} + \f^{*} Z^{(G+3)} \frac{ }{ } \right)  + \ 
\rm{terms} \  \rm{independent} \  \rm{of} \  \rm{the} \  \rm{antifields}.
\label{CohomolAFD6}
\eq
where  $X^{(G+1)\m\n}$, $X^{(G+2)\m}$, $Y^{(G+2)\m}$, $X^{(G+3)}$, 
$Y^{(G+3)}$ 
and $Z^{(G+3)}$ are arbitrary polynomials independent of the
 external fields.

Solving the condition of $\cs_{\S}^0$-invariance of 
(\ref{CohomolAFD6}), we can show, in a similar way as in $D=2$, that the 
dependence on the external fields is $\cs_{\S}^0$-trivial.

Once the antifields have been eliminated, the cohomology may 
depend on the remaining fields which  do not belong to doublets. 
In particular, {\it at the non-integrated level}, the cohomology
will depend only on  $\pa_{[\r}A_{\m\n]}$, $\pa_{[\m}\f_{\n]}$, 
$\f$, $\pa_{(\m}\a_{\n)}$\,\footnote{ $\pa_{(\m}\a_{\n)}$ 
denotes symmetric derivative in $\m$, $\n$}, 
their derivatives, and $\ome$.

The cocycles independent of the external fields will be 
denoted as 
\be
\D^{(G)}=\int d^6x \ Q^{(G)}(x)\ .
\ee
The condition of $\cs_{\S}^0$-invariance leads to the following set
 of descent 
equations: 
\bq
&&\cs_{\S}^0 Q^{(G+k)}_{[\m_{1}..\m_{k}]} = 
\pa^{\l} Q^{(G+k+1)}_{[\m_{1}..\m_{k}\l]}
\label{DesEqD6}
\eq
with $k=0,..,6$ and, since the dimension of the space-time 
is equal to 6, 
$Q^{(G+7)}_{[\m_{1}..\m_{7}]}=0$.

The dimension and ghost number of the fields forbid any possible 
non-trivial solution in the sector $G=1$, implying the absence of 
anomaly in the theory.

In the sector $G=0$, the most general solution for the top level of 
the descent equations reads 
\bq
&&Q^{(0)} = \Ome(\pa_{[\r}A_{\m\n]},\bar{a}),
\label{Cohomols0D6}
\eq
 where $\Ome(\pa_{[\r}A_{\m\n]},\bar{a})$ is a polynomial depending 
on $\pa_{[\r}A_{\m\n]}$, 
with coefficients as arbitrary funcions of $\bar{a}$.

The condition of invariance under the full operator $\cs_{\S}$, 
the constraints (\ref{eqmotaD6}) and (\ref{constraintD6})
being taken into account, leads to
\bq
&&\D^{(0)} = \S_{\rm{inv}}\ ,
\label{CohomolsD6}
\eq
where $\S_{\rm{inv}}$ is the invariant action \equ{invD6}. 
However, the latter is a trivial cocycle:
\bq
\S_{\rm{inv}}\!\!\!&=&\!\!\! \frac{1}{2} \ \cs_{\S} \int d^6x \left(
\frac{1}{2}A_{\m\n}^{*}A^{\m\n} - \a^{*}_{\m}\a^{\m} - \f^{*}_{\m}\f^{\m} +
\a^{*}\a + \ome^{*}\ome + \f^{*}\f \right.\nonumber \\
&&\left.-\bar\a^{\m}\pa^{\n}A_{\m\n} -  \bar\f^{\m}A_{\m\n}\pa^{\n}a -
\bar{\f}^\prime \pa^{\m}a \f_{\m} -\bar \a \pa^{\m}\a_{\m} - 
\bar\f\pa^{\m}\f_{\m} \frac{}{}\right). \label{StrivD6}
\eq
As a conclusion, the PST model in $D=6$ is finite up to 
non-physical renormalizations.
\section{The PST model in D=4}

A $p$-form gauge potential, with non-vanishing $(p+1)$-form 
self-dual field strength in a Minkowski space-time, can only 
exist in dimensions $D=2(p+1)$ for $p$ even, what restricts the 
relevant dimensions to $2,6,10,...$

In order to cross over this restriction and define a 
covariant duality symmetric action in $D=4$, the PST 
recipe~\cite{Pasti} 
is to introduce two gauge potentials $A_{\m}^{a}$, $a=1,2$, 
the self-duality condition being now a duality relation 
between the two field strengths $F_{\m\n}^{a}$ of the theory.

The PST action in $D=4$ is
\bq
\S_{\rm{inv}}\!\!\!&=&\!\!\! \int d^4x \ 
\left(-\frac{1}{8}F^{a\m\n} F_{\m\n}^{a} 
- \frac{1}{4(\pa a)^2}\ 
\pa^{\m}a \ \cf_{\m\n}^{a} \cf^{a\n\r} \pa_{\r}a \right)
\label{invD4}
\eq
where $F_{\m\n}^{a} \equiv \pa_{\m}A_{\n}^{a} - \pa_{\n}A_{\m}^{a}$ is 
the field strength of the gauge potential $A_{\m}^{a}$, $a$ is 
a scalar auxiliary field 
and 
\be
\cf_{\m\n}^{a} \equiv \cl^{ab}F_{\m\n}^{b}
 - \frac{1}{2} \e_{\m\n\r\s}F^{a\r\s} = \frac{1}{2} 
\e_{\m\n\r\s}\cl^{ab}\cf^{b\r\s}\ ,
\ee
where $\cl$ is the antisymmetric ($2\times 2$)-matrix with 
$\cl^{12}=1$.

The action (\ref{invD4}) has the following symmetries 
(we only display the non-zero variations):
\bq
&&\d_{I}A_{\m}^a = \pa_{\m} \a^a\ , \nonumber \label{invID4} \\
&&\d_{II}A_{\m}^a = \f^a \pa_{\m}a\ ,\nonumber \label{invIID4} \\
&&\d_{III}A_{\m}^a = \b \hat{B}^{a}_{\m}\ ,\quad
\d_{III} a = \b\ ,\quad {\rm{with}}\,\,B^{a}_{\m} 
\equiv \frac{\cf^{a}_{\m\n}\pa^{\n}a}{(\pa a)^2}\ 
{\rm{and}}\ \hat{B}^{a}_{\m}=\cl^{ab}
B^{b}_{\m}\ ,\nonumber \label{invIIID4} \\
&&\d_{IV}A_{\m}^a = f^{a}_{\m}(a)\ ,\quad{\rm{with}}\ 
f^{a}_{\m}(a)\ \mbox{an arbitrary function of}\ a\ .\label{invIVD4}
\eq
We again must require $\pa^{\m}a\pa_{\m}a \neq 0$ and so split $a(x)$ 
in $\bar a(x)$ 
and $a^\prime(x)$ according to (\ref{splitD2}).

The set of local symmetries in (\ref{invIVD4}) leads to the following BRST 
transformations :
\bq
&&sA_{\m}^a = \pa_{\m} \a^a + \f^a \pa_{\m}a + \b \hat{B}^{a}_{\m},
~~\nonumber\\
&&sa^\prime = \b,~~\nonumber\\
&&s\b=0~~,\nonumber\\
&&s\a^a = -\b\f^a~~,\nonumber\\
&&s\f^a = \frac{\b}{(\pa a)^2}(\pa^{\m}a\pa_{\m}\f^{a} - \hat{B}^{a}_{\m}\pa^{\m}\b)~~,
\label{BRSTD4} 
\eq
where $\a^a$,$\f^{a}$ and $\b$ are now Faddeev-Popov ghosts.

Following the same steps as before, we write down
the total action as 
\bq
\S = \S_{\rm{inv}} + \S_{\rm{gf}} + \S_{\rm{ext}},
\label{totactionD4}
\eq
with $\S_{\rm{inv}}$ given in (\ref{invD4}) and 
\bq
&&\S_{\rm{gf}} = s \int d^4x \ (\bar{\a}^a \pa^{\m} A_{\m}^a + 
\bar{\f}^a A_{\m}^a \pa^{\m}a +
\bar{\b} a^\prime), \nonumber \\
&&\S_{\rm{ext}} = \int d^4x \ ( A_{\m}^{*a} sA^{a\m} + 
\a^{*a} s\a^a + \f^{*a} s\f^a ) ,
\label{gfextActionsD4}
\eq   
where we have introduced the external fields coupled to the non-linear 
BRST transformations, as well as  
antighosts and Lagrange multipliers, transforming as
\bq 
&&s\,{\bar{C}} = \Pi \ , \quad s\,\Pi = 0~, 
\label{BRSTAntigLagD4}  
\eq 
with $\bar{C} = (\bar\a^{a},\,\bar\f^{a},\,\b)$ and $\Pi = 
(\p^{a},\,\r^{a},\,b)$. 

The dimension and ghost number of all fields are displayed below in 
Table 5.
\begin{table}[hbt]
\centering
\begin{tabular}{|c||c|c|c|c|c|c|c|c|c|c|c|c|c|c|c|}
\hline
\phantom{$\dfrac{x}{x}$}
& $A_{\m}^a$ & $a^\prime$ & $\b$ & $\a^a$ & $\f^a$ & $\bar{\b}$ & $b$ & 
$\bar{\a}^{a}$ & $\p^a$ & $\bar{\f}^a$ & $\r^a$ & $A_{\m}^{*a}$ & 
$\a^{*a}$ & $\f^{*a}$ & 
$s$ \\ \hline\hline
$d$ & $1$ & $-1$ & $-1$ & $0$ & $1$ & $5$ & $5$ & $2$ & $2$ & 
$3$ & $3$ & $3$ & $4$ & $3$ & $0$\\ 
\hline
$\F\Pi$ & $0$ & $0$ & $1$ & $1$ & $1$ & $-1$ & $0$ & $-1$ & $0$ & $-1$
 & $0$ & $-1$ & $-2$ & $-2$ 
& $1$ \\ \hline
\end{tabular}
\caption[t1]{dimension $d$ and ghost number $\F\Pi$.}
\label{table5}
\end{table} 

The Slavnov-Taylor identity obeyed by the total classical action  
(\ref{totactionD4}) 
is found to be
\bq
{\cal S}(\S)\!\!\!&=&\!\!\! \int {d^4x}\left({\frac{\d\S}{\d A_{\m}^{*a}}
\frac{\d\S}{\d A^{a\m}}}+{\frac{\d\S}{\d\a^{*a}}\frac{\d\S}{\d\a^{a}}}+
{\frac{\d\S}{\d\f^{*a}}\frac{\d\S}{\d\f^{a}}}~+\right. \nonumber \\
&&+\left.
{\b\frac{\d\S}{\d a^{\prime}}}+{\p^{a}\frac{\d\S}{\d \bar{\a}^{a}}}+
\r^a{\frac{\d\S}{\d \bar{\f}^{a}}}+b{\frac{\d\S}{\d\bar\b}} \right)=0,~~
\label{slavnovD4} 
\eq
which leads to the following nilpotent linearized Slavnov-Taylor operator :
\bq
{\cs_{\S}}\!\!\!&=&\!\!\! \int {d^4x}\left(\frac{\d\S}{\d
A_{\m}^{*a}}\frac{\d}{\d A^{a\m}}+ \frac{\d\S}{\d
A^{a\m}}\frac{\d}{\d A_{\m}^{*a}}+ \frac{\d\S}{\d\a^{*a}}
\frac{\d}{\d\a^a}+
\frac{\d\S}{\d\a^{a}}\frac{\d}{\d\a^{*a}}~+\right. \nonumber \\
&&+\left.\frac{\d\S}{\d\f^{*a}}\frac{\d}{\d\f^a}+\frac{\d\S}{\d\f^a}
\frac{\d}{\d\f^{*a}}+
\p^{a}\frac{\d}{\d \bar{a}^a}+\r^a \frac{\d}{\d\bar{\f}^{a}}+
b \frac{\d}{\d\bar{\b}}\right). ~~
\label{LinSlavD4}
\eq
The total classical action (\ref{totactionD4}) obeys again 
strong constraints, which can be 
extended to the quantum level and are important in the study of 
the cohomology of $\cs_{\S}$. 
The integrated antighost equation for $\a$, the integrated equation 
of motion for $a^\prime$ and the 
constraint between $a^\prime$ and $\bar{a}$ are respectively given by  
\bq
&&\int d^4x \ \frac{\d\S}{\d\a^{a}} = 0 , \nonumber \\
&&\int d^4x \ \frac{\d\S}{\d a^{\prime}} = \int d^4x \ b, \nonumber \\
&&\left({\frac{\d}{\d a^\prime}}-{\frac{\d}{\d\bar{a}}}\right)\S = b.
\label{constraintsD4}
\eq
The cohomology of $\cs_{\S}$ can be suitably analyzed by expanding it 
first according to a counting operator like (\ref{countopD2}), acting 
on all fields,  
and then by expanding the lowest order contribution of this series in 
powers 
of $\f^a$ 
only. Following exactly the same procedure as in the cases $D=2$ and
 $D=6$ 
the cohomology 
of the lowest contribution of the last expansion (in powers of $\f^a$) is 
shown to be empty in the sector with ghost number $1$, implying the absence 
of anomaly. For ghost number 0 we find, at lowest order, the quadratic part 
of the invariant action \equ{invD4}, which corresponds to the full
invariant action for the exact $\SS_\S$ operator. However, this term
again is $\SS_\S$-trivial:
\bq
\S_{\rm{inv}}\!\!\!&=&\!\!\! \frac{1}{2} \ \cs_{\S} \int d^4x \left(
A_{\m}^{*a}A^{a\m} - \a^{*a}\a^{a} - \f^{*a}\f^{a} 
-\bar{\a}^a \pa^{\m} A_{\m}^a -\bar{\f}^a A_{\m}^a \pa^{\m}a  \right) \ , 
\label{StrivD4}
\eq
hence the model is finite.
\section{Conclusions}
We have shown, by a purely algebraic analysis independent of any 
regularization scheme, 
the absence of gauge anomaly and 
the finiteness of the PST model -- up to possible singularities 
absorvable by local field redefinitions -- in two, four and six dimensions. 
Both results were established by proving that the BRST cohomology 
is trivial in the sectors with ghost numbers $0$ and $1$. We believe 
it is possible to extend 
this proof to all 
dimensions for which the PST model can be defined. This is a topic  
we still intend to exploit.

 Another point worth to  be noticed
concerns the implementation of the 
self-duality constraint at the quantum level. As said before it 
can be obtained 
at the classical approximation by means of 
the equation of motion of $\f$ and the global 
symmetry present in the PST model -- e.g. given by the transformation
\equ{trans2D2}, in the 2-dimensional case. 
However, the implemention of this argument at 
the quantum regime is still an open problem. In particular, the validity 
of the global symmetry has not been 
shown yet. This kind of discussion remains for future work.

\subsubsection*{Acknowledgements} 
We would like to thank Alexandre de Mello Delpupo for his participation 
to an early stage of this work.  Olivier Piguet is very grateful to the
DCP of the Centro Brasileiro de Pesquisas F\'\i sicas (CBPF), as well as 
to the Universidade Cat\'olica de Petr\'opolis, for various
invitations during which part important of this work was done.
Oswaldo M. Del Cima 
would like to express his gratitude to 
J.A. Helay\"el-Neto and his colleagues at GFT-UCP for the warm
hospitality and friendship. He dedicates this work to 
his wife, Zilda Cristina, to his daughter, Vittoria, and to his son,
Enzo. 

\end{document}